\documentclass[10pt,aps,prb,twocolumn]{revtex4}   

\usepackage{graphicx}

\usepackage{amsmath}    
\usepackage{graphicx}   
\usepackage{color}

\newcommand{\sovast}{Soviet~Ast.}      

\renewcommand{\vec}[1]{\mathbf{#1}}
\newcommand{\vecs}[1]{\boldsymbol{#1}}
\newcommand{\unit}[1]{\hat{\vec{#1}}}
\newcommand{\units}[1]{\hat{\vecs{#1}}}

\def\del{\nabla}

\def\cross{\times}
\def\real{\operatorname{Re}}

\def\be{\begin{equation}}
\def\ee{\end{equation}}
\def\bi{\begin{itemize}}
\def\ei{\end{itemize}}
\def\ben{\begin{enumerate}}
\def\een{\end{enumerate}}

\begin{document}

\title{Understanding the gravitational-wave 
Hellings and Downs curve for \\
pulsar timing arrays
in terms of sound and electromagnetic waves}
\author{Fredrick A.\ Jenet}
 \email{merlyn@alum.mit.edu}   
 \affiliation{University of Texas at Brownsville, 
 Department of Physics and Astronomy, and
 Center for Advanced Radio Astronomy,
 Brownsville, TX 78520}
\author{Joseph D.\ Romano}
 \email{joseph.romano@utb.edu}   
 \affiliation{University of Texas at Brownsville, 
 Department of Physics and Astronomy, and
 Center for Gravitational-Wave Astronomy,
 Brownsville, TX 78520}

\date{\today}

\begin{abstract}
Searches for stochastic gravitational-wave backgrounds using pulsar
timing arrays look for correlations in the timing residuals
induced by the background across the pulsars in the array.
The correlation signature of an isotropic, unpolarized
gravitational-wave background 
predicted by general relativity 
follows the so-called {\em Hellings and Downs} curve, 
which is a relatively simple function of the angle between
a pair of Earth-pulsar baselines.
In this paper, we give a pedagogical discussion of the 
Hellings and Downs curve for pulsar timing arrays, considering 
simpler analogous scenarios involving sound and electromagnetic waves.
We calculate Hellings-and-Downs-type functions for these
two scenarios and develop a framework suitable 
for doing more general correlation calculations. 
\end{abstract}

\maketitle

\section{Introduction}

\begin{figure*}
\begin{center}
\includegraphics[angle=0, width=.9\textwidth]{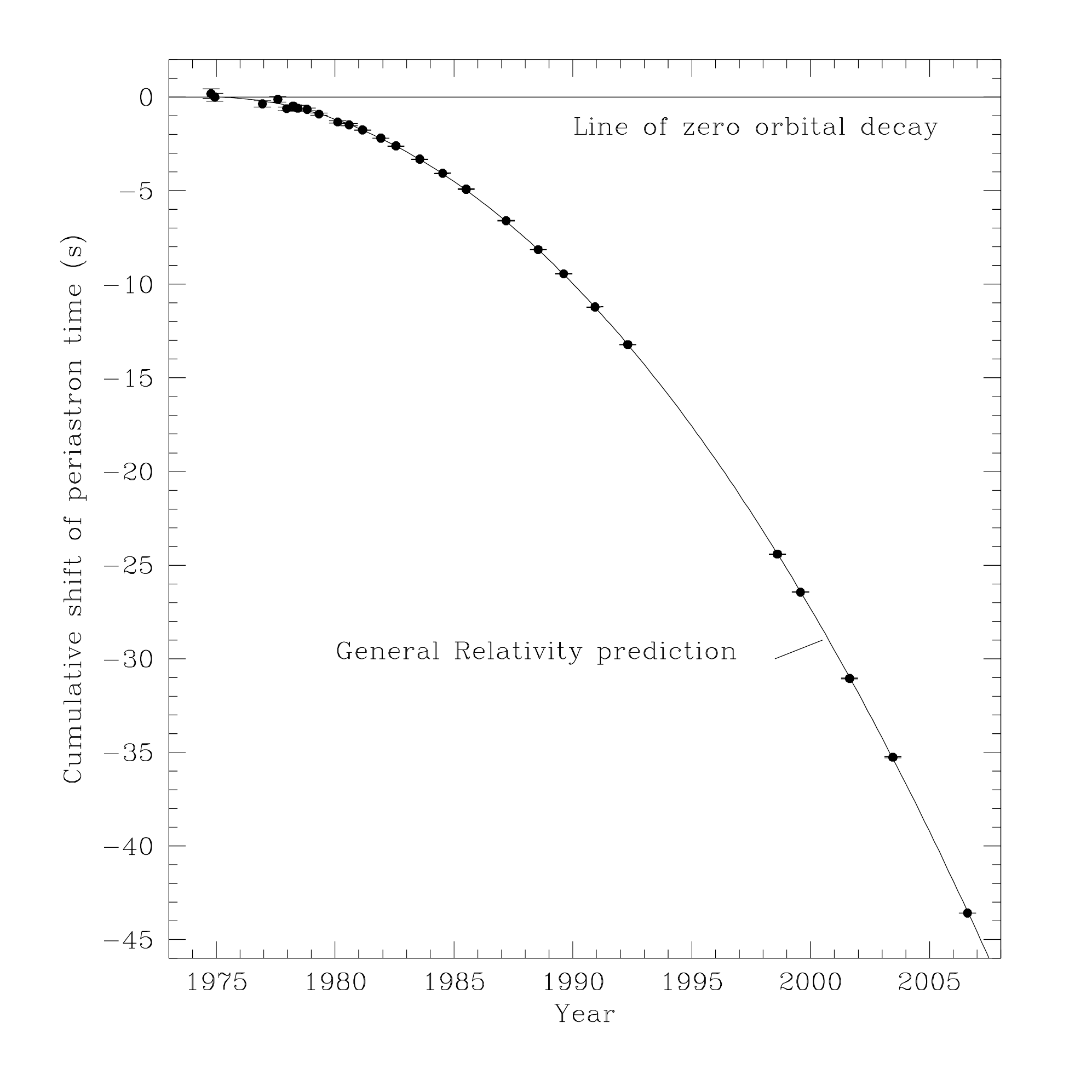}
\caption
{Decrease in the orbital period of binary pulsar 
PSR~B1913+16.\cite{weisberg-et-al:2010}
The measured data points and error bars agree with the 
prediction of general relativity (parabola) for the rate of orbital 
decay due to gravitational-wave emission.}
\label{f:Weisberg-et-al}
\end{center}
\end{figure*}

A pulsar is a rapidly-rotating neutron star that emits a 
beam of electromagnetic radiation (usually in the form of 
radio waves) from its magnetic poles.\cite{lorimer-kramer:2004}
If the beam of radiation crosses our line of sight, we  
see a flash of radiation, similar to that of a lighthouse beacon.
These flashes can be thought of as ticks of a giant astronomical clock,
whose regularity rivals that of the best human-made atomic clocks.
By precisely monitoring the arrival times of the pulses, 
astronomers can determine:
(i) intrinisic properties of the pulsar---e.g., its rotational
period and whether it is spinning up or spinning down;
(ii) extrinsic properties of the pulsar---e.g., whether it 
is in a binary system, and if so what are its orbital 
parameters; and 
(iii) properties of the intervening `stuff' between us 
and the pulsar---e.g., the column density of electrons in
the interstellar medium.\cite{lorimer-LRR:2008, stairs-LRR:2003}
Indeed, it was the precise monitoring (for over 30 years)
of the pulses from binary pulsar PSR~B1913+16 that has 
given us the most compelling evidence to date for the
existence of gravitational waves.\cite{weisberg-et-al:2010}
The measured decrease in the orbital period of binary pulsar 
PSR~B1913+16 agrees precisely with the predictions of general 
relativity for the energy loss due to gravitational-wave 
emission.
(See Fig.~\ref{f:Weisberg-et-al}.)
This was a path-breaking result, with the discovery of the 
binary pulsar\cite{Hulse-Taylor:1975} being worthy of a 
Nobel Prize in Physics for Joseph Taylor and Russell Hulse in 1993.

Monitoring the gravitational-wave-induced decay of a binary 
system, like PSR~B1913+16, is one method for detecting gravitational waves.
Another method is to look for the effect of gravitational 
waves on the radio pulses that propagate from a pulsar to a 
radio antenna on Earth.
The basic idea is that 
when a gravitational wave transits the Earth-pulsar 
line of sight, it creates a perturbation in the intervening 
spatial metric, causing a change in the propagation 
time of the radio pulses emitted by the 
pulsar.\cite{estabrook-wahlquist:1975,sazhin:1978,detweiler:1979}
(This is the timing {\em response} of the Earth-pulsar baseline 
to a gravitational wave.)
One can then compare the measured and predicted 
times of arrival (TOAs) of the pulses, using timing 
models that take into account the various intrinsic 
and extrinsic properties of the pulsar.
Since standard timing models factor in only deterministic 
influences on the arrival times of the pulses, 
the difference between the measured and predicted
TOAs will result in a stream of {\em timing residuals}, 
which encode the influence of both deterministic and
stochastic (i.e., random) gravitational waves 
as well as any other random noise processes on the measurement.\cite{footnote1}
If one has a set of radio pulsars, i.e., a pulsar 
timing array (PTA),
one can correlate 
the residuals across pairs of Earth-pulsar baselines, 
leveraging the common 
influence of a background of gravitational waves against 
unwanted, uncorrelated noise.
The key property of a PTA is that the signal from a 
stochastic gravitational-wave background 
will be correlated across the baselines, while that from the 
other noise processes will not.
This is what makes a PTA function as a galactic-scale,
gravitational-wave detector.\cite{Foster-Backer:1990}

\begin{figure*}
\begin{center}
\includegraphics[angle=0, width=.7\textwidth]{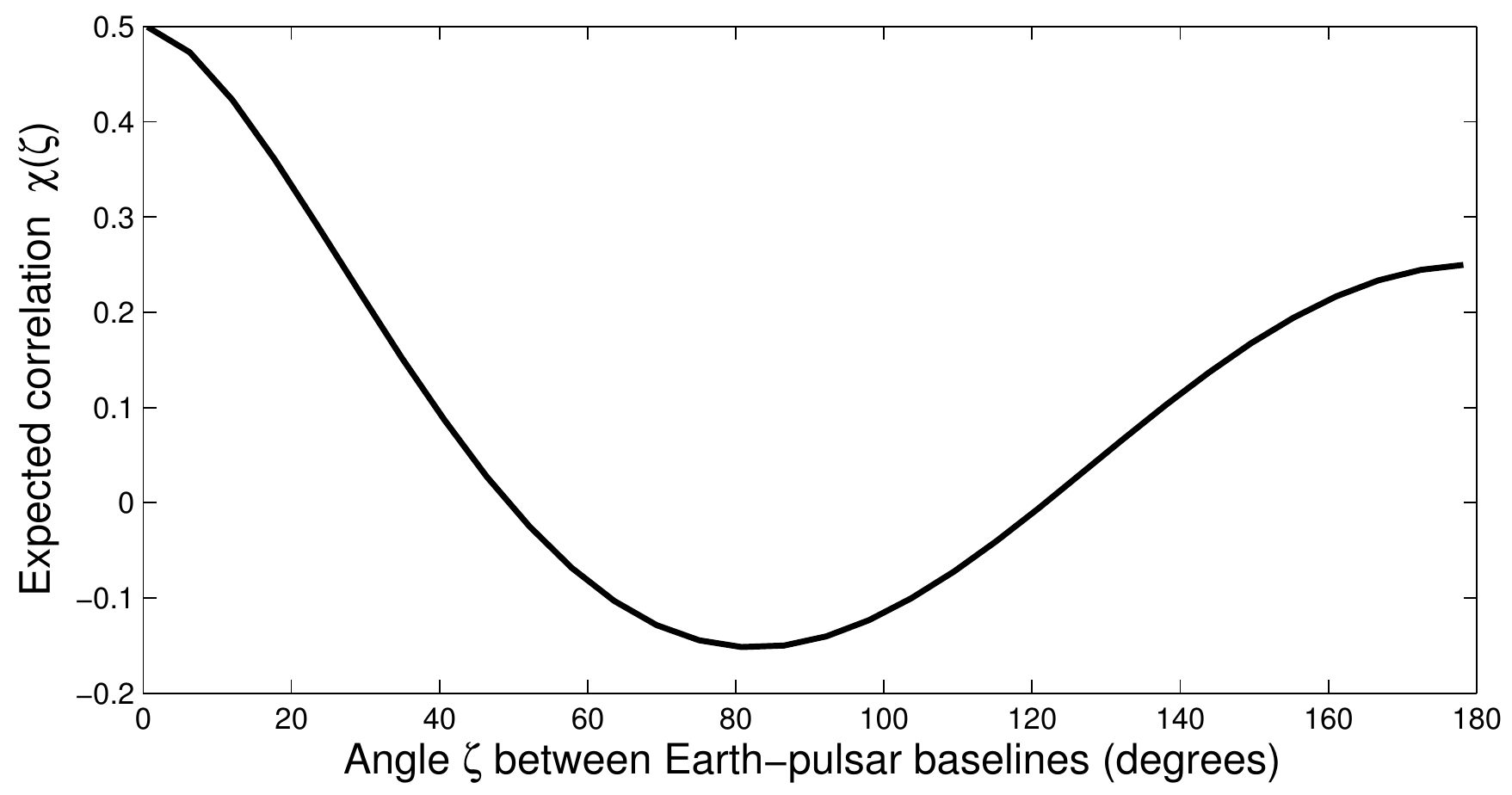}
\caption
{Hellings and Downs curve for the expected correlated response
of a pair of Earth-pulsar baselines to an isotropic, unpolarized
stochastic gravitational-wave background, plotted as a function of the angle 
between the baselines,
cf.~Eq.~(\ref{e:HDcurve}).}
\label{f:HDcurve}
\end{center}
\end{figure*}

For an isotropic, unpolarized stochastic background of quadrupole
gravitational radiation composed of the plus ($+$) and cross ($\cross$) 
polarization modes predicted by general relativity, 
the expected correlated response of a pair of 
Earth-pulsar baselines to the
background follows the so-called {\em Hellings and Downs} curve,
named after the two authors who first calculated it in 1983.\cite{Hellings-Downs:1983}
A plot of the Hellings and Downs curve as a function of the
angle between a pair of baselines is shown in Fig.~\ref{f:HDcurve}.
Searches for stochastic gravitational-wave backgrounds using
pulsar timing arrays effectively compare the measured 
correlations with the expected values from the 
Hellings and Downs curve 
to determine whether or not a signal from an isotropic,
unpolarized background is present (or absent) in the data.
Gravitational-wave backgrounds predicted by alternative 
theories of gravity, which have different polarization 
modes,\cite{Lee-et-al:2008}
or backgrounds that have an anisotropic distribution of 
gravitational-wave energy 
on the sky,\cite{Mingarelli-et-al:2013, Taylor-Gair:2013, Gair-et-al:2014}
will induce different correlation signatures and must be searched
for accordingly.
To date no detections have been made,
but upper limits on the strength of the background 
have been set\cite{Shannon-et-al:2013} that constrain 
certain models of gravitational-wave backgrounds produced by 
the inspirals of binary supermassive black holes (SMBHs) in merging 
galaxies throughout the universe.

\begin{figure*}
\begin{center}
\includegraphics[angle=0, width=.6\textwidth]{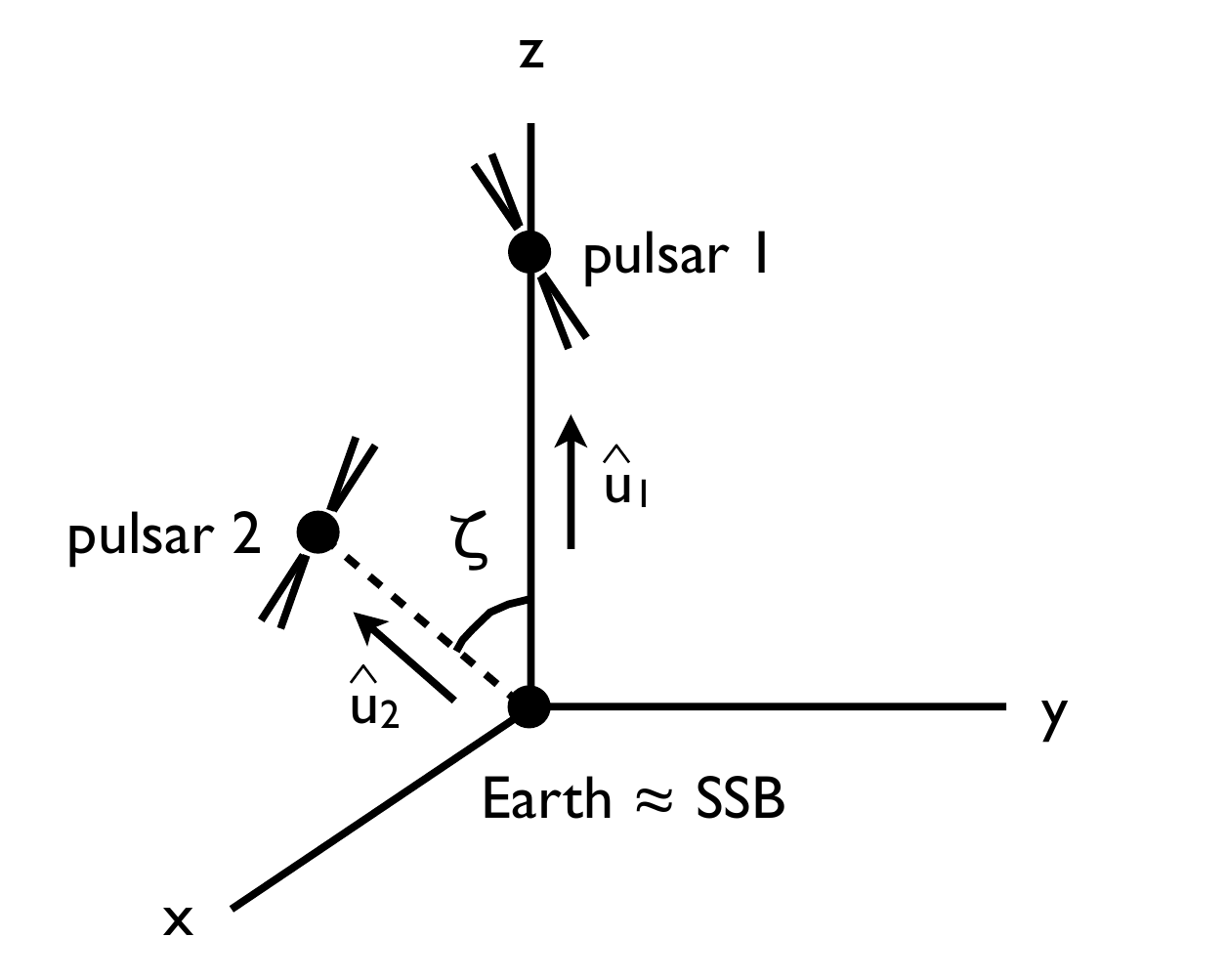}
\caption
{Geometry for the calculation of the Hellings and Downs
function for the correlated response of a pair of 
Earth-pulsar baselines to an isotropic, unpolarized 
stochastic gravitational-wave background.
The coordinate system is chosen so that the Earth 
is located at the origin and pulsar 1 is located on the 
$z$-axis, a distance $D_1$ from the origin.
Pulsar 2 is located in the $xz$-plane, 
a distance $D_2$ from the origin. 
The two Earth-pulsar baselines point along the 
unit vectors $\unit u{}_1$ and $\unit u{}_2$.
The angle between the two baselines is denoted by $\zeta$,
and is given by $\cos\zeta = \unit u{}_1\cdot \unit u_{2}$.
(Actually, the origin of the coordinate system for the
calculation is the fixed solar system barycenter (SSB), 
and not the moving Earth.
But since the the distance between the Earth and SSB 
(1~au) is much smaller than the typical distance 
to the two pulsars, 
$D_{1,2}\sim 1~{\rm kpc}=2\times 10^8~{\rm au}$,
there is no practical difference between the Earth-pulsar 
and SSB-pulsar baselines.)
}
\label{f:pulsar-geom}
\end{center}
\end{figure*}

Mathematically, the Hellings and Downs curve is the sky-averaged 
and polarization-averaged product of the response of a pair of 
Earth-pulsar baselines 
to a plane wave propagating in a particular direction 
with either $+$ or $\cross$ polarization.
It has the analytic form
\be
\chi(\zeta)=
\frac{1}{2}
-\frac{1}{4}\left(\frac{1-\cos\zeta}{2}\right)
+\frac{3}{2}\left(\frac{1-\cos\zeta}{2}\right)
\ln\left(\frac{1-\cos\zeta}{2}\right)\,,
\label{e:HDcurve}
\ee
where $\zeta$ is the angle between two Earth-pulsar baselines.\cite{footnote2}
(See Fig.~\ref{f:pulsar-geom} for the Earth-pulsar baseline geometry.)
The integration that one must do in order to obtain the
above expression is non-trivial enough that 
Hellings and Downs originally used the symbolic manipulation computer
system MACSYMA to do the calculation.\cite{Hellings-Downs:1983}
It turns out that is also possible to evaluate the 
integral by hand,
using contour integration for part of the integration
(see, e.g., Appendix~\ref{s:app}).
But for some reason, perhaps related to the difficulty of 
analytically evaluating the sky integral, students or beginning
researchers who are first 
introduced to the Hellings and Downs curve see it as a somewhat 
mysterious object, intimately connected to the realm of pulsar timing.
Granted, the precise analytic form in (\ref{e:HDcurve}) is 
specific to the response of a pair of Earth-pulsar baselines 
to an isotropic, unpolarized stochastic gravitational-wave background,
but Hellings-and-Downs-type functions show up in {\em any} scenario where
one is interested in the dependence of the correlated 
response of a pair of receivers on the geometrical configuration
of the two receivers.
The geometry relating the configuration of one receiver to 
another might be more complicated (or simpler) than that for 
the pulsar timing case, but the basic idea of correlation 
across receivers is exactly the same.

The purpose of this paper is to emphasize this commonality, and to
calculate Hellings-and-Downs-type functions 
for two simpler scenarios.
Scenario 1 will be for a pair of receivers constructed from 
omni-directional microphones responding to an isotropic stochastic
sound field.
Scenario 2 will be for a pair of receivers constructed from 
electric dipole antennas responding to an isotropic and unpolarized
stochastic electromagnetic field.
These two scenarios were chosen since the derivation of the 
corresponding Hellings-and-Downs-type functions 
(cf.~Eqs.~(\ref{e:Gamma12_sound}) and (\ref{e:Gamma12_EM})) and the 
evaluation of the necessary sky-integral and polarization averaging 
(for the electromagnetic-wave case) are relatively simple.
But the steps that one must go through to obtain these results are 
identical to those for the gravitational-wave pulsar timing 
Hellings and Downs function, even though the mathematics needed to
derive the relevant expression for the pulsar timing case 
(cf.~Eq.~(\ref{e:HD_pulsartiming})) is more involved.
Hopefully, after reading this paper, 
the reader will understand the pulsar timing Hellings and Downs 
curve in its proper context, and appreciate that it is a special 
case of a general correlation calculation.

The rest of the paper is organized as follows:
In Section~\ref{s:random}, we describe a general mathematical
formalism for working with random fields, which we will use 
repeatedly in the following sections.
In Section~\ref{s:sound} we apply this formalism to
calculate a Hellings-and-Downs-type function for the case 
of omni-directional microphones 
in an isotropic stochastic sound field.
Section~\ref{s:EM} extends the calculation to electric 
dipole antennas in an isotropic and unpolarized stochastic
electromagnetic field, which requires us to deal with 
the polarization of the component waves.
Finally, in Section~\ref{s:summary}, we summarize the 
basic steps needed to calculate Hellings-and-Downs-type
functions in general, and then set-up up the calculation 
for the actual pulsar timing Hellings and Downs curve, 
leaving the evaluation of the final integral to the motivated 
reader.
(We have included details of the calculation in Appendix~\ref{s:app}, 
in case the reader has difficulty completing the calculation.)

\section{Random fields and expectation values}
\label{s:random}

Probably the most important reason for calculating Hellings-and-Downs-type 
functions is to determine the correlation signature of a signal buried
in  noisy data.
The situation is tricky when the signal is associated 
with a {\em random} field 
(e.g., for a stochatic gravitational-wave background),
since then one is effectively trying to detect ``noise in noise."
Fortunately, it turns out that there is a way to surmount this problem.
The key idea is that although the signal associated with a random field
is typically indistinguishable from noise in a {\em single} detector or 
receiving system, it is correlated between {\em pairs} of detectors 
or receiving systems
in ways that differ, in general, from instrumental or measurement noise.
(In other words, by using multiple detectors, one can leverage the common 
influence of the background field against unwanted, 
uncorrelated noise processes.)
At each instant of time, the measured correlation is simply the product 
of the output of two detectors.
But since both the field and the instrumental noise are random processes,
the measured correlation will fluctuate with time as dictated by the 
statistical properties of the field and noise.
By averaging the correlations over time, we obtain an estimate of the 
{\em expected value} of the correlation, which we can then compare with 
predicted values assuming the presence (or absence) of a signal.
The purpose of this section is to develop the mathematical machinery
that will allow us to perform these statistical correlation calculations.

In the following three sections of the paper, we will be working with fields 
(sound, electromagnetic, and gravitational fields) that
are made up of waves propagating in all different directions.
These waves, having been produced by a large number of 
independent and uncorrelated sources, will have, 
in general, different frequencies, amplitudes, and phases.
(In the case of electromagnetic and gravitational waves, 
they will also have different polarizations.)
Such a superposition of waves is most conveniently 
described {\em statistically}, in terms of a Fourier 
integral whose Fourier coefficients are {\em random} variables.
The statistical properties of the field will then
be encoded in the statistical properities of the 
Fourier coefficients, which are much simpler to work
with as we shall show below, cf.~Eqs.~(\ref{e:A})
and (\ref{e:AA}).

To illustrate these ideas as simply as possible, we 
will do the calculations in this section for 
an arbitrary {\em scalar} field $\Phi(t,\vec x)$. 
Analogous calculations would also go through 
for vector and tensor fields 
(e.g., electromagnetic and gravitational fields) 
with mostly just an increase in notational complexity,
coming from the vector and tensor nature of these
fields and their polarization properties.
Scalar fields are particularly simple since they are 
described by a single real (or complex) number at each 
point in space $\vec x$, at each instant of time $t$.
Sound waves, which we will discuss in detail in 
Section~\ref{s:sound}, are an example of a scalar 
field.
The Fourier integral for a scalar field $\Phi(t,\vec x)$
has the form
\be
\tilde \Phi(t, \vec x)
= \int {\rm d}^3\vec k\> \tilde 
A(\vec k)e^{i(\vec k\cdot \vec x -\omega t)}\,,
\quad
\Phi(t,\vec x) = \real[\tilde \Phi(t,\vec x)]\,,
\label{e:Phitilde}
\ee
with $\omega/k = v$, where $v$ is the speed of 
wave propagation and $k=|\vec k|$.
The relation $\omega/k = v$ is required for 
$e^{i(\vec k\cdot\vec x-\omega t)}$ to be a 
solution of the wave equation.
The Fourier coefficients $\tilde A(\vec k)$ are
complex-valued random variables, 
and can be written as
\be
\tilde A(\vec k) = A(\vec k)e^{i\alpha(\vec k)}
=a(\vec k) + i b(\vec k)\,,
\label{e:ab}
\ee
where $A$, $\alpha$, $a$, and $b$ are all real-valued
functions of $\vec k$.

The statistical properties of the field $\Phi(t,\vec x)$
are completely determined by the joint probability distributions
\be
p_n(\Phi_1,t_1,\vec x_1;\Phi_2,t_2,\vec x_2; \cdots ;\Phi_n,t_n,\vec x_n)\,,
\quad
n=1,2,\cdots
\label{e:p_n}
\ee
in terms of which one can calculate 
the expectation values 
\be
\langle \Phi(t_1,\vec x_1)\rangle\,,\quad
\langle \Phi(t_1,\vec x_1)\Phi(t_2, \vec x_2)\rangle\,,\quad
{\rm etc.}
\label{e:EVs}
\ee
For example, the expectation value of the field at spatial
location $\vec x_1$ at time $t_1$ is defined by
\be
\langle \Phi(t_1,\vec x_1)\rangle
\equiv
\int_{-\infty}^\infty {\rm d}\Phi_1\>
\Phi_1(t_1,\vec x_1)p_1(\Phi_1,t_1,\vec x_1)\,.
\ee
Equivalently, the expectation values can be defined in terms of
an {\em ensemble} average, e.g.,
\be
\left\langle \Phi(t, \vec x)\right\rangle
\equiv \lim_{N\rightarrow\infty}
\frac{1}{N}\sum_{i=1}^{N} 
\Phi^{(i)}(t, \vec x)\,,
\ee
where $\Phi^{(i)}(t, \vec x)$ denotes a particular realization of 
$\Phi(t,\vec x)$.
The usefulness of knowing the expectation values given in 
(\ref{e:EVs}) is that such knowledge is equivalent to knowing 
the joint probability distributions (\ref{e:p_n}) and hence
the complete statistical properties of the field.\cite{footnote3}
These expectation values 
in turn are completely encoded in the expectation
values of the products of the Fourier coefficients 
$\tilde A(\vec k)$.

The simplest case, which is also the one we consider, 
is for a multivariate {\em Gaussian}-distributed field, 
since knowledge of the quadratic expectation values 
is sufficient to determine all higher-order moments.
Without loss of generality, we will work with the real 
random variables $a(\vec k)$ and $b(\vec k)$, and assume 
that any non-zero constant value has been subtracted from
the field:
\be
\langle a(\vec k)\rangle = 0\,,
\quad
\langle b(\vec k)\rangle = 0\,.
\label{e:ab=0}
\ee
We will also assume that the 
field $\Phi(\vec x,t)$ is {\em stationary in time} and 
{\em spatially homogeneous}---i.e., that the statistical
properties of the field are unaffected by a change in
either the origin of time or the origin of spatial coordinates:
\be
t\rightarrow t+t'\,,
\quad
\vec x\rightarrow \vec x+\vec x'\,.
\ee
This means that the quadratic expectation values
can depend only on the {\em difference} between these
coordinates:
\be
\langle \Phi(t_1,\vec x_1)\Phi(t_2,\vec x_2)\rangle 
= C(t_1-t_2,\vec x_1-\vec x_2)\,.
\label{e:stationary-homog}
\ee
Such behavior follows from
\be
\begin{aligned}
\langle a(\vec k)a(\vec k')\rangle &= \frac{1}{2}B(\vec k)\,\delta^3(\vec k-\vec k')\,,
\\
\langle b(\vec k)b(\vec k')\rangle &= \frac{1}{2}B(\vec k)\,\delta^3(\vec k-\vec k')\,,
\\
\langle a(\vec k)b(\vec k')\rangle &= 0\,,
\label{e:abEVs}
\end{aligned}
\ee
with
\begin{multline}
C(t_1-t_2,\vec x_1-\vec x_2) = \frac{1}{2}\int {\rm d}^3\vec k\>
B(\vec k)
\\
\times\cos\left[\vec k\cdot(\vec x_1-\vec x_2)-\omega(t_1-t_2)\right].
\label{e:C-B}
\end{multline}
For readers interested in proving this last statement, write $\Phi$ 
as the sum $\Phi = (\tilde\Phi+\tilde\Phi^*)/2$ and then use 
Eqs.~(\ref{e:Phitilde}) and (\ref{e:ab}) to expand the 
left-hand-side of Eq.~(\ref{e:stationary-homog}) in terms of 
expectation values of $a(\vec k)$ and $b(\vec k)$.
Given Eq.~(\ref{e:abEVs}), Eq.~(\ref{e:stationary-homog}) then 
follows with $C(t_1-t_2,\vec x_1-\vec x_2)$ given by Eq.~(\ref{e:C-B}).

The physical meaning of the Dirac delta functions that appear in 
the expectation values of Eq.~(\ref{e:abEVs}) is that waves propagating in 
different directions $\unit k$ and $\unit k'$
and having different angular frequencies $\omega=kv$ and $\omega'=k'v$
are {\em statistically independent} of one another.
In other words, the expected correlations are non-zero only for waves 
traveling in the same direction and having the same frequency.
Using Eqs.~(\ref{e:ab=0}) and (\ref{e:abEVs}), it is also 
straightforward to show that the complex Fourier coefficients 
$\tilde A(\vec k) = a(\vec k) +ib(\vec k)$ satisfy
\be
\langle \tilde A(\vec k)\rangle = 0\,,
\quad
\langle \tilde A^*(\vec k)\rangle = 0\,,
\label{e:A}
\ee
and that
\be
\begin{aligned}
\langle \tilde A(\vec k)\tilde A(\vec k')\rangle &= 0\,,
\\
\langle \tilde A^*(\vec k)\tilde A^*(\vec k')\rangle &= 0\,,
\\
\langle \tilde A(\vec k)\tilde A^*(\vec k')\rangle &= 
B(\vec k)\,\delta^3(\vec k-\vec k')\,.
\label{e:AA}
\end{aligned}
\ee
These two sets of expectation values for the
Fourier coefficients $\tilde A(\vec k)$ are the main 
results of this section.
The vanishing of the first two expectation values
in (\ref{e:AA}) imply
\begin{widetext}
\be
\begin{aligned}
\left\langle \Phi^2(t, \vec x)\right\rangle
&= \frac{1}{4}\left\langle (\tilde \Phi(t,\vec x)+\tilde\Phi^*(t, \vec x))
(\tilde\Phi(t,\vec x)+\tilde\Phi^*(t, \vec x))\right\rangle
\\
&= \frac{1}{4}\left\{
\langle\tilde\Phi(t,\vec x)\tilde\Phi(t,\vec x)\rangle
+\langle\tilde\Phi^*(t, \vec x)\tilde\Phi^*(t, \vec x)\rangle
+\langle\tilde\Phi(t,\vec x)\tilde\Phi^*(t, \vec x)\rangle
+\langle\tilde\Phi^*(t, \vec x)\tilde\Phi(t, \vec x)\rangle
\right\}
\\
&= \frac{1}{2}\left\langle \tilde\Phi(t,\vec x)\tilde\Phi^*(t, \vec x)\right\rangle\,,
\end{aligned}
\ee
which we will use repeatedly in the following sections.
\end{widetext}

As discussed at the start of this section, we are 
ultimately interested in calculating the expected
correlation $\langle r_1(t)r_2(t)\rangle$ of the responses 
$r_1(t)$, $r_2(t)$ of two receiving systems 
$R_1$, $R_2$ to the field $\Phi(t,\vec x)$.
It is this expected correlation that we can compare 
against the actual measured correlation,
assuming that the other noise processes are uncorrelated
across different receiving systems.
The response of the receiving systems will be linear 
in the field, given by a 
convolution of $R_1$ and $R_2$ with $\Phi$.
Since $\Phi(t,\vec x)$ is a random field, $r_1(t)$ and $r_2(t)$ 
will be random functions of time.
In addition, the expectation value 
$\langle r_1(t_1) r_2(t_2)\rangle$ 
will depend only on the time difference $t_1-t_2$,
as a consequence of our assumption regarding 
stationarity of the field, cf.~(\ref{e:stationary-homog}).
Hence, the expected correlation 
$\langle r_1(t)r_2(t)\rangle$ 
will be {\em independent} of time, and we expect to be able
to estimate this correlation by averaging 
together measurements made at different instants of time:
\be
\langle r_1(t) r_2(t)\rangle 
\equiv \lim_{N\rightarrow\infty}\frac{1}{N} \sum_{i=1}^{N}
r_1(t_i) r_2(t_i)\,.
\ee
Random processes for which this is true---i.e., for which
time averages equal ensemble averages over different 
realizations of the field---are said to be {\em ergodic}.

In what follows we will assume that all our random processes are 
ergodic so that ensemble averages can be replaced
by time averages (and/or spatial averages) if desired.
This will allow us to calculate expectation values by averaging
over segments of a {\em single} realization, which is usually 
all that we have in practice.
Although ergodicity is often a good assumption to make, 
it is important to note that 
not all stationary random processes are ergodic.
An example\cite{helstrom:1968} of a stationary 
random process that is 
not ergodic is an ensemble of constant time-series $x(t)=a$, 
where the values of $a$ are uniformly distributed between 
$-1$ and $1$.
The esemble average $\langle x(t)\rangle =0$ for all $t$, 
but the time-average of a single realization equals the 
value of $a$ for whichever time-series is drawn from the ensemble.
For simplicity of presentation in the remainder of this paper, 
we will continue to treat the Fourier expansion coefficients 
as random variables and calculate ensemble averages of these 
quantities, rather than time (and/or spatial) averages of products 
of the plane wave components $e^{i(\vec k\cdot \vec x -\omega t)}$.

\section{Scenario 1: Sound waves}
\label{s:sound}

The first scenario we consider involves sound.
Mathematically, sound waves in air are pressure 
deviations (relative to  atmospheric pressure)
that satisfy the 3-dimensional wave equation.
If we denote the pressure deviation at time $t$ and
spatial location $\vec x$ by $p(t, \vec x)$, then
\be
\del^2 p-\frac{1}{c_s^2}
\frac{\partial^2 p}{\partial t^2}
=0\,,
\label{e:sound-wave-eqn}
\ee
where $\del^2$ denotes the Laplacian\cite{footnote4}
and $c_s$ denotes the speed of sound in air
(approximately $340~{\rm m/s}$ at room temperature).
The most general solution of the 3-dimensional wave
equation is a superposition of plane waves:
\be
p(t,\vec x) = 
\int{\rm d}^3\vec k\> 
A(\vec k) \cos(\vec{k}\cdot\vec x-\omega t + \alpha(\vec k))\,,
\ee
where the wave vector $\vec k$ and angular frequency $\omega$ 
are related by $\omega/k = c_s$
in order that (\ref{e:sound-wave-eqn}) be satisfied for each $\vec k$.
As discussed in Section~\ref{s:random}, 
it will be more convienent 
to work with the complex-valued solution
\be
\tilde p(t,\vec x) = 
\int{\rm d}^3\vec k\> 
\tilde A(\vec k)e^{i(\vec{k}\cdot\vec x-\omega t)},
\quad
\tilde A(\vec k) = A(\vec k)e^{i\alpha(\vec k)}\,,
\label{e:ptilde}
\ee
for which $p(t,\vec x)$ is the real part.

For a stochastic sound field, the Fourier coefficients $\tilde A(\vec k)$
are random variables.
We will assume that these coefficients satisfy (\ref{e:A}) and (\ref{e:AA}), 
with the additional requirement that the function
$B(\vec k)$ be independent of direction $\unit k$, which is
appropriate for a statistically {\em isotropic} sound field.
(This means there is no preferred direction of wave propagation
at any point in the field.)
As we shall now show, the function $B(\vec k) \equiv B(k)$ is 
simply related to the power per unit frequency in the sound 
field integrated over all directions.
To prove this last claim we calculate the mean-squared pressure
deviations:
\begin{widetext}
\be
\begin{aligned}
\langle p^2(t, \vec x)\rangle
&=\frac{1}{2}\langle \tilde p(t,\vec x)\tilde p^*(t, \vec x)\rangle
\\
&= \frac{1}{2}\left\langle
\int {\rm d}^3\vec k\>
\tilde A(\vec k)\,e^{i(\vec k\cdot\vec x-\omega t)}
\int {\rm d}^3\vec k'\>
\tilde A^*(\vec k')\,e^{-i(\vec k'\cdot\vec x-\omega' t)}
\right\rangle
\\
&=
\frac{1}{2}
\int {\rm d}^3\vec k\>\int {\rm d}^3\vec k'\>
\left\langle \tilde A(\vec k)\tilde A^*(\vec k')\right\rangle
\,
e^{i(\vec k-\vec k')\cdot\vec x}
e^{-i(\omega-\omega') t}
\\
&=
\frac{1}{2}
\int {\rm d}^3\vec k\>B(k)
\\
&=
\frac{1}{2}
\int_0^\infty k^2{\rm d}k\>
\int_{S^2}{\rm d}^2\Omega_{\unit k}\>
B(k)
\\
&=2\pi
\int_0^\infty k^2{\rm d}k\>
B(k)\,.
\end{aligned}
\ee
\end{widetext}
Thus, if we write
\be
\langle p^2(t,\vec x)\rangle 
=\int_0^\infty {\rm d}\omega\>
\frac{{\rm d}\langle p^2\rangle}{{\rm d}\omega}\,,
\ee
then 
\be
\frac{{\rm d}\langle p^2\rangle}{{\rm d}\omega}
=\frac{2\pi k^2}{c_s} B(k)
\label{e:powerSound}
\ee
as claimed.
%
%

\begin{figure*}
\begin{center}
\includegraphics[angle=0, width=.6\textwidth]{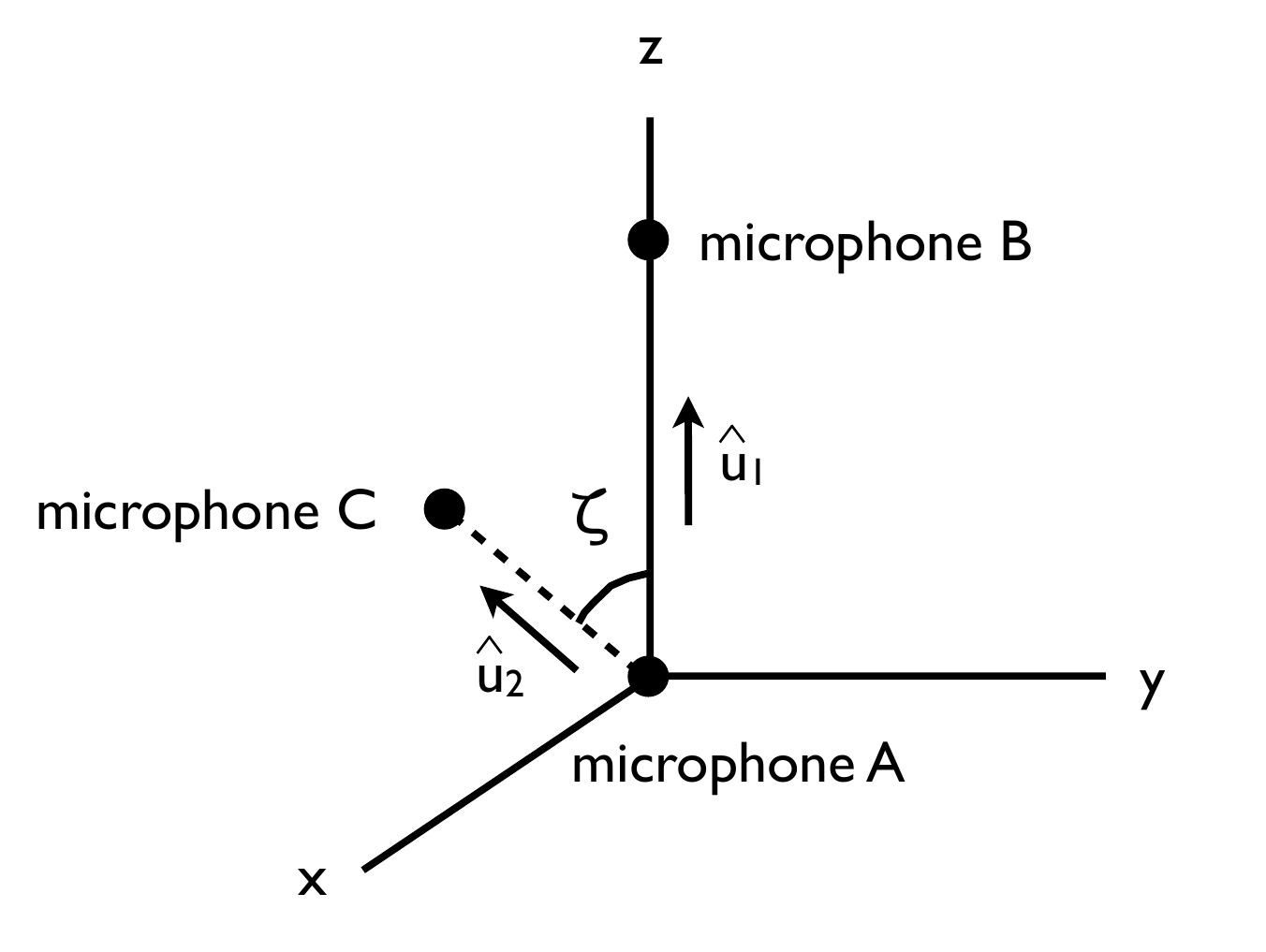}
\caption
{Geometry for the calculation of the Hellings-and-Downs-type 
function for a pair of receivers constructed from 
omni-directional microphones responding to an isotropic stochastic 
sound field.
Receiving system 1 is constructed from microphones $A$ and $B$,
and points along the unit vector $\unit u{}_1$.
Receiving system 2 is constructed from microphones $A$ and $C$,
and points along the unit vector $\unit u{}_2$.
The coordinate system is chosen so that microphone $A$, 
which is common to both recieving systems, is located 
at the origin.
Microphone $B$ is located on the $z$-axis, a distance $D_B$ 
from the origin, while microphone $C$ is located in the 
$xz$-plane, a distance $D_C$ from the origin.  
The angle between the two receiving systems is denoted by $\zeta$,
and is given by $\cos\zeta = \unit u{}_1\cdot \unit u_{2}$.
(Note the similarity of this figure and Fig.~\ref{f:pulsar-geom}.)}
\label{f:sound-geom}
\end{center}
\end{figure*}

To determine the acoustical analogue of the pulsar timing
Hellings and Downs function, we need to calculate the expected 
correlation of the responses $r_1(t)$ and $r_2(t)$ of two 
receiving systems to an isotropic stochastic sound field.
A single receiving system will consist of a {\em pair} of 
omni-directional (i.e., isotropic) microphones that are 
separated in space as shown in Fig.~\ref{f:sound-geom}.
For simplicity, we will assume that the microphones are
indentical and have a gain $G$ that is independent of frequency.
The response  $r_1(t)$ of receiving system 1, 
consisting of microphones $A$ and $B$, is defined to be
the real part of 
\be
\tilde r_1(t) = \tilde V_A(t)-\tilde V_B(t)\,,
\ee
where 
\be
\tilde V_A(t) 
= G\,\tilde p(\vec x_A,t)\,,
\quad
\tilde V_B(t) 
= G\,\tilde p(\vec x_B,t)\,.
\ee
The response of receiving system 2, consisting of microphones
$A$ and $C$, is defined similarly,
\be
\tilde r_2(t) = \tilde V_A(t)-\tilde V_C(t)\,,
\ee
with microphone $C$ replacing microphone $B$.
Note that microphone $A$ is common to both receiving systems,
and that we have taken the time of the measurement to be the same
at both microphones, which physically corresponds to running 
equal-length wires from each microphone to our receiving system.

\begin{figure}
\begin{center}
\includegraphics[angle=0, width=.49\textwidth]{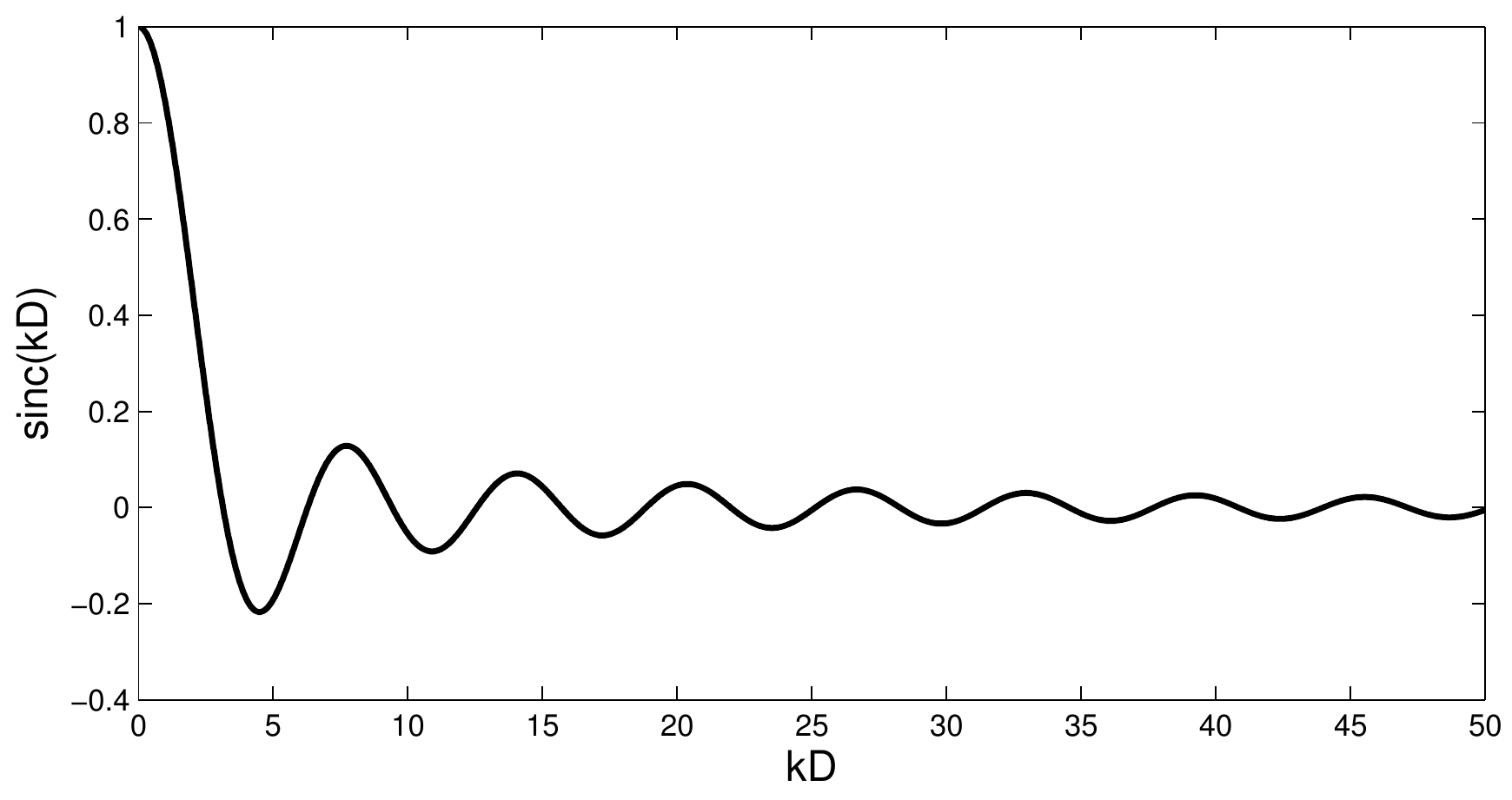}
\caption
{Plot of ${\rm sinc}(kD)$ versus $kD$.}
\label{f:sinc_curve}
\end{center}
\end{figure}

The expected value of the correlated response is then
\begin{widetext}
\be
\begin{aligned}
\langle r_1(t)r_2(t)\rangle
&=\frac{1}{2}
{\rm Re}\left\{
\langle \tilde r_1(t)\tilde r_2^*(t)\rangle
\right\}
\\
&= 
\frac{1}{2}
{\rm Re}\left\{
\left\langle 
\left(\tilde V_A(t)-\tilde V_B(t)\right)
\left(\tilde V^*_A(t)-\tilde V^*_C(t)\right)
\right\rangle
\right\}
\\
&=
\frac{1}{2}
{\rm Re}\left\{
G^2\int {\rm d}^3\vec k\>\int {\rm d}^3\vec k'\>
\left\langle \tilde A(\vec k)\tilde A^*(\vec k')\right\rangle
e^{-i(\omega-\omega') t}
\left[1-e^{i\vec k\cdot\vec x_B}\right]
\left[1-e^{-i\vec k'\cdot\vec x_C}\right]
\right\}
\\
&=
\frac{1}{2}
{\rm Re}\left\{
G^2\int {\rm d}^3\vec k\>
B(k)
\left[1-e^{i\vec k\cdot\vec x_B}\right]
\left[1-e^{-i\vec k\cdot\vec x_C}\right]
\right\}
\\
&=
\frac{1}{2}
{\rm Re}\left\{
G^2\int_0^\infty k^2{\rm d}k\> B(k)
\int_{S^2}{\rm d}^2\Omega_{\unit k}\>
\left[1-e^{i\vec k\cdot\vec x_B}
-e^{-i\vec k\cdot\vec x_C} + e^{i\vec k\cdot(\vec x_B-\vec x_C)}\right]
\right\}
\\
&=
\int_0^\infty {\rm d}\omega\> 
\frac{{\rm d}\langle p^2\rangle}{{\rm d}\omega}\,
\Gamma_{12}(\omega)\,,
\label{e:Czeta_sound}
\end{aligned}
\ee
where the correlation function
\be
\Gamma_{12}(\omega) 
= 
{\rm Re}\left\{
G^2\,
\frac{1}{4\pi}\int_{S^2}{\rm d}^2\Omega_{\unit k}\>
\left[1-e^{i\vec k\cdot\vec x_B}
-e^{-i\vec k\cdot\vec x_C} + e^{i\vec k\cdot(\vec x_B-\vec x_C)}\right]
\right\}\,.
\label{e:Gamma12_sound}
\ee
\end{widetext}
The integrals of the exponentials over all directions $\unit k$ 
are of the form
$\int_{S^2}{\rm d}^2\Omega_{\unit k}\>e^{i\vec k\cdot\vec x}$,
where $\vec x$ is a fixed vector.
Such an integral is most easily evaluated in a frame in which 
the $z$-axis is directed along $\vec x$.
In this frame,
\be
\begin{aligned}
\int_{S^2}{\rm d}^2\Omega_{\unit k}\>e^{i\vec k\cdot\vec x}
&=\int_0^{2\pi}{\rm d}\phi\>\int_{-1}^1{\rm d}(\cos\theta)\>
e^{ikD\cos\theta}
\\
&=2\pi \frac{1}{ikD}\left(e^{ikD}-e^{-ikD}\right)
\\
&=4\pi\,{\rm sinc}(kD)\,,
\end{aligned}
\ee
where $D=|\vec x|$ and ${\rm sinc}(x) \equiv \sin x/x$.
Since the sinc function rapidly approaches zero for $x\gg 1$
as shown in Fig.~\ref{f:sinc_curve},
we can ignore the contribution from the last three integrals
in (\ref{e:Gamma12_sound}) provided $kD\gg 1$, or 
equivalently, provided $D\gg 1/k = c_s/\omega$.
This is called the {\em short-wavelength approximation}.
For audible sound, which has frequencies $f\equiv \omega/(2\pi)$ 
in the range $\sim 20~{\rm Hz}$ to $\sim 20~{\rm kHz}$, 
this condition becomes
\be
D\gg \frac{c_s}{\omega} = \frac{340~{\rm m/s}}{2\pi\cdot 20~{\rm Hz}}
=2.7~{\rm m}\,.
\ee
So assuming that the individual microphones are separated by
more than this amount, we have
\be
\chi(\zeta)\equiv
\Gamma_{12}(\omega) \simeq G^2\,.
\label{e:Gamma12_sound_approx}
\ee
In other words, 
the Hellings and Downs function for an isotropic stochastic
sound field is simply a constant, 
independent of the angle between the two receiving systems.
The expected correlation is thus
\be
\langle r_1 r_2\rangle
\simeq G^2\,\langle p^2\rangle\,,
\ee
which is the mean power in the sound field multiplied
by a constant $G^2$.
This result is to be expected for omni-directional microphones 
in an isotropic stochastic sound field.

Although this was a somewhat long calculation to obtain an 
answer that, in retrospect, did not require {\em any} calculation, 
the formalism developed here can be applied 
with rather minor modifications to handle more complicated 
scenarios as we shall see below.

\section{Scenario 2: Electromagnetic waves}
\label{s:EM}

The second example we consider involves electromagnetic waves.
Similar to sound, electromagnetic waves are solutions to a 
3-dimensional wave equation, but with the speed of light 
$c=2.998\times 10^8$ replacing the speed of sound $c_s$:
\be
\begin{aligned}
\del^2 \vec E -\frac{1}{c^2}\frac{\partial^2 \vec E}{\partial t^2}
&=0\,,
\\
\del^2 \vec B -\frac{1}{c^2}\frac{\partial^2 \vec B}{\partial t^2}
&=0\,.
\end{aligned}
\ee
The most general solution to the wave equation for the electric 
and magnetic fields is given by a sum of plane waves
similar to that in Eq.~(\ref{e:Phitilde}),
\be
\begin{aligned}
&\tilde{\vec E}(t, \vec x)
=\int {\rm d}^3\vec k\>\left\{
\tilde E_{1}(\vec k)
\units\epsilon_{1}(\unit k)
+\tilde E_{2}(\vec k)
\units\epsilon_{2}(\unit k)
\right\}
e^{i(\vec k\cdot\vec x-\omega t)}\,,
\\
&\tilde{\vec B}(t, \vec x) 
=\int {\rm d}^3\vec k\>\frac{\unit k}{c}\cross 
\left\{\tilde E_{1}(\vec k)
\units\epsilon_{1}(\unit k)
+\tilde E_{2}(\vec k)
\units\epsilon_{2}(\unit k)
\right\}
e^{i(\vec k\cdot\vec x-\omega t)}\,,
\label{e:EB1}
\end{aligned}
\ee
with 
\be
\vec E(t, \vec x) = \real[\tilde{\vec E}(t, \vec x)]\,,
\quad
\vec B(t, \vec x) = \real[\tilde{\vec B}(t, \vec x)]\,,
\label{e:EB2}
\ee
and $\omega/k = c$.
In the above expressions,
$\units\epsilon_{\alpha}(\hat k)$ $(\alpha = 1,2)$
are two unit polarization vectors, orthogonal to one another and to the 
direction of propagation:
\be
\units\epsilon_{\alpha}(\unit k)\cdot\units\epsilon_{\beta}(\unit k)=\delta_{\alpha\beta}\,,
\quad
\unit k\cdot\units\epsilon_{\alpha}(\unit k)=0\,.
\ee
Note that there is freedom to rotate the polarization vectors 
in the plane orthogonal to $\unit k$.
For simplicity, we will choose 
\be
\begin{aligned}
\units\epsilon_{1}(\unit k)
&=\cos\theta\cos\phi\,\unit x
+\cos\theta\sin\phi\,\unit y
-\sin\theta\,\unit z
=\units\theta\,,
\\
\units\epsilon_{2}(\unit k)
&=-\sin\phi\,\unit x + \cos\phi\,\unit y
=\units\phi\,,
\label{e:eps12}
\end{aligned}
\ee
whenever
$\unit k$ points in the direction given by
the standard angular coordinates $(\theta,\phi)$ on the sphere:
\be
\unit k = \sin\theta\cos\phi\,\unit x
+\sin\theta\sin\phi\,\unit y
+\cos\theta\,\unit z\,.
\ee
Since the receiving systems that we shall consider below are
constructed from electric dipole antennas, which respond only 
to the electric part of the field, we will ignore the magnetic
field for the remainder of this discussion.

For a stochastic field, the Fourier coefficients are complex-valued
random variables.  
We will assume that they have expectation values 
(cf.~(\ref{e:A}) and (\ref{e:AA})):
\be
\langle\tilde E_{\alpha}(\vec k)\rangle=0\,,
\quad
\langle\tilde E^*_{\alpha}(\vec k)\rangle=0\,,
\label{e:E}
\ee
and
\be
\begin{aligned}
\langle\tilde E_{\alpha}(\vec k)\tilde E_{\beta}(\vec k')\rangle
&=0\,,
\\
\langle\tilde E^*_{\alpha}(\vec k)\tilde E^*_{\beta}(\vec k')\rangle
&=0\,,
\\
\langle\tilde E_{\alpha}(\vec k)\tilde E^*_{\beta}(\vec k')\rangle
&=\delta^3(\vec k-\vec k')\,P_{\alpha\beta}(\vec k)\,.
\end{aligned}
\label{e:EE}
\ee
As before, the Dirac delta function ensures that the
radiation propagating in different directions and having
different angular frequencies are statistically independent of 
one another.
If the field is also statistically 
isotropic and unpolarized, then the polarization 
tensor $P_{\alpha\beta}(\vec k)$ will be proportional to the identity matrix 
$\delta_{\alpha\beta}$, with a proportionality constant 
independent of direction on the sky:
\be
P_{\alpha\beta}(\vec k)
=P(k)\,\delta_{\alpha\beta}\,.
\ee
Similar to the case for sound, the function $P(k)$ turns out to be 
simply related to the power per unit frequency in 
the electric field when summed over both polarization modes and 
integrated over all directions.
To see this we calculate mean-squared electric field:
\begin{widetext}
\be
\begin{aligned}
\langle E^2(t, \vec x) \rangle
& = \frac{1}{2}\langle \tilde{\vec E}(t, \vec x)\cdot \tilde{\vec E}^*(t, \vec x)\rangle
\\
&= \frac{1}{2}\left\langle
\int {\rm d}^3\vec k\>\sum_{\alpha=1,2}
\tilde E_{\alpha}(\vec k)\,\hat{\vecs\epsilon}_{\alpha}(\unit k)\,e^{i(\vec k\cdot\vec x-\omega t)}
\cdot
\int {\rm d}^3\vec k'\>\sum_{\beta=1,2}
\tilde E^*_{\beta}(\vec k')\,\hat{\vecs\epsilon}_{\beta}(\unit k')\,e^{-i(\vec k'\cdot\vec x-\omega' t)}
\right\rangle
\\
&=
\frac{1}{2}
\int {\rm d}^3\vec k\>\int {\rm d}^3\vec k'\>
\sum_{\alpha=1,2}\sum_{\beta=1,2}
\left\langle \tilde E_{\alpha}(\vec k)\tilde E^*_{\beta}(\vec k')\right\rangle
\hat{\vecs\epsilon}_{\alpha}(\unit k)\cdot\hat{\vecs\epsilon}_{\beta}(\unit k')
\,
e^{i(\vec k-\vec k')\cdot\vec x}
e^{-i(\omega-\omega') t}
\\
&=\frac{1}{2}\int {\rm d}^3\vec k\>
P(k)
\sum_{\alpha=1,2}
\hat{\vecs\epsilon}_{\alpha}(\unit k)\cdot\hat{\vecs\epsilon}_{\alpha}(\unit k)
\\
&=4\pi \int_0^\infty k^2{\rm d}k\>P(k)
=\int_0^\infty {\rm d}\omega\> \frac{{\rm d}\langle E^2\rangle}{{\rm d}\omega}\,,
\end{aligned}
\ee
\end{widetext}
for which
\be
\frac{{\rm d}\langle E^2\rangle}{{\rm d}\omega}
= \frac{4\pi k^2}{c}P(k)
\label{e:powerEM}
\ee
as claimed.
Note that this has the same form as that for sound,
cf.~(\ref{e:powerSound}), with the speed 
of light $c$ replacing the speed of sound $c_s$,
and the extra factor of two coming from the summation
over the two (assumed statistically equivalent) 
polarization modes for the electromagnetic field.
%

\begin{figure*}
\begin{center}
\includegraphics[angle=0, width=.6\textwidth]{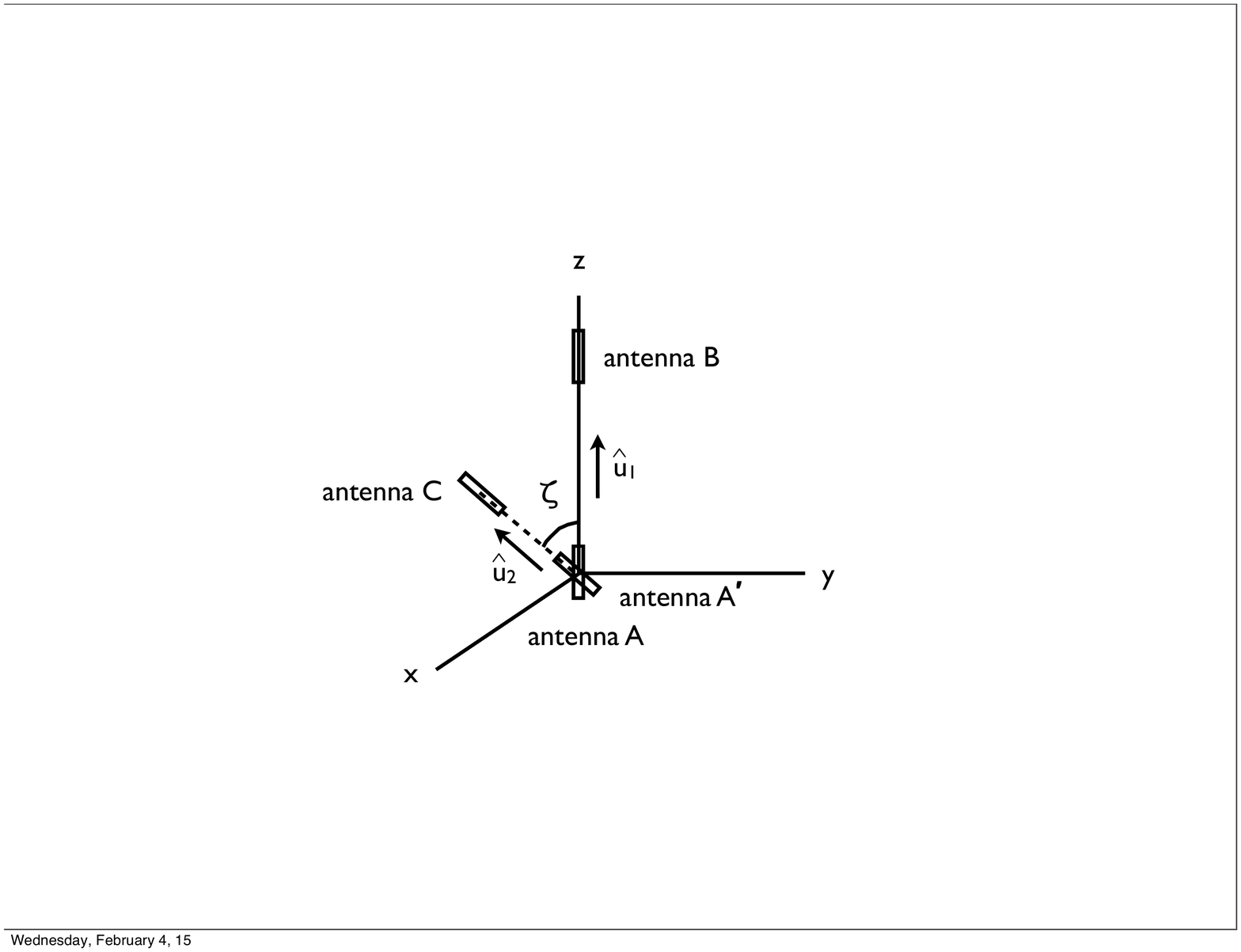}
\caption
{Geometry for the calculation of the Hellings-and-Downs-type 
function for a pair of receivers constructed from 
electric dipole antennas responding to an isotropic, unpolarized
stochastic electromagnetic field.
Receiving system 1 is constructed from dipole antennas 
$A$ and $B$, which are both directed along $\unit u_1$, 
which points from $A$ to $B$.
Receiving system 2 is constructed from dipole antennas 
$A'$ and $C$, which are both directed along $\unit u_2$, 
which points from $A'$ to $C$.
The coordinate system is chosen so that the two dipole antennas 
$A$ and $A'$ are located at the origin.
Dipole antenna $B$ is located on the $z$-axis, 
a distance $D_B$ from the origin, while 
dipole antenna $C$ is located in the $xz$-plane, 
a distance $D_C$ from the origin.
The angle between the two receiving systems is denoted by $\zeta$,
and is given by $\cos\zeta = \unit u{}_1\cdot \unit u_{2}$.
(Again, note the similarity of this figure and Fig.~\ref{f:pulsar-geom}.)}
\label{f:EM-geom}
\end{center}
\end{figure*}

To determine the electromagnetic analogue of the pulsar timing Hellings
and Downs function, we need to calculate the expected correlation of 
the responses $r_1(t)$ and $r_2(t)$ 
of two receiving systems to an
isotropic, unpolarized stochastic electromagnetic field.
A single receiving system will consist of a pair of electric dipole 
antennas that are separated in space as shown in 
Fig.~\ref{f:EM-geom}.
For simplicity, we will assume that the electric dipole antennas 
are identical and short relative to the wavelengths that make 
up the electric field.
The response $r_1(t)$ of receiving system 1, consisting of 
electric dipole antennas $A$ and $B$, is defined to be the
real part of
\be
\tilde r_1(t) = \tilde V_A(t) - \tilde V_B(t)\,,
\ee
where
\be
\tilde V_A(t) 
= \unit u_1\cdot \tilde{\vec E}(\vec x_A,t)\,,
\quad
\tilde V_B(t) 
= \unit u_1\cdot \tilde{\vec E}(\vec x_B,t)\,.
\ee
The response $r_2(t)$ of receiving system 2, consisting of electric dipole
antennas $A'$ and $C$, is defined similarly
\be
\tilde r_2(t) = \tilde V_{A'}(t) -  \tilde V_C(t)\,,
\ee
where 
\be
\tilde V_{A'}(t) 
= \unit u_2\cdot \tilde{\vec E}(\vec x_A,t)\,,
\quad
\tilde V_C(t) 
= \unit u_2\cdot \tilde{\vec E}(\vec x_C,t)\,.
\ee
Note that $\tilde V_{A'}(t)$ differs from $\tilde V_A(t)$ since 
the dipole antenna for $A'$ points along $\unit u_2$, while that
for $A$ points along $\unit u_1$.

The expected value of the correlated response is then
\begin{widetext}
\be
\begin{aligned}
\langle r_1(t)r_2(t)\rangle
&=\frac{1}{2}
{\rm Re}\left\{
\langle \tilde r_1(t)\tilde r_2^*(t)\rangle
\right\}
\\
&= 
\frac{1}{2}
{\rm Re}\left\{
\left\langle 
\left(\tilde V_A(t)-\tilde V_B(t)\right)
\left(\tilde V^*_{A'}(t)-\tilde V^*_C(t)\right)
\right\rangle
\right\}
\\
&=
\frac{1}{2}
{\rm Re}\Bigg\{
\int {\rm d}^3\vec k\>\int {\rm d}^3\vec k'\>
\sum_{\alpha=1,2}\sum_{\beta=1,2}
\left\langle \tilde E_{\alpha}(\vec k)\tilde E^*_{\beta}(\vec k')\right\rangle
\unit u_1\cdot\hat{\vecs\epsilon}_{\alpha}(\unit k)\,
\unit u_2\cdot\hat{\vecs\epsilon}_{\beta}(\unit k')
\\
&\hspace{2.5in}
\times
e^{-i(\omega-\omega') t}
\left[1-e^{i\vec k\cdot\vec x_B}\right]
\left[1-e^{-i\vec k'\cdot\vec x_C}\right]
\Bigg\}
\\
&=
\frac{1}{2}
{\rm Re}\left\{
\int {\rm d}^3\vec k\>
P(k)
\sum_{\alpha=1,2}
(\unit u_1\cdot\hat{\vecs\epsilon}_{\alpha}(\unit k))
(\unit u_2\cdot\hat{\vecs\epsilon}_{\alpha}(\unit k))
\left[1-e^{i\vec k\cdot\vec x_B}\right]
\left[1-e^{-i\vec k\cdot\vec x_C}\right]
\right\}
\\
&=
\frac{1}{2}
{\rm Re}\left\{
\int_0^\infty k^2{\rm d}k\>P(k)
\int_{S^2} {\rm d}\Omega_{\unit k}\>
\sum_{\alpha=1,2}
(\unit u_1\cdot\hat{\vecs\epsilon}_{\alpha}(\unit k))
(\unit u_2\cdot\hat{\vecs\epsilon}_{\alpha}(\unit k))
\left[1-e^{i\vec k\cdot\vec x_B}\right]
\left[1-e^{-i\vec k\cdot\vec x_C}\right]
\right\}
\\
&=
\int_0^\infty {\rm d}\omega\>
\frac{{\rm d}\langle E^2\rangle}{{\rm d}{\omega}}\,
\Gamma_{12}(\omega)\,,
\end{aligned}
\ee
where 
\be
\Gamma_{12}(\omega) = 
{\rm Re}\left\{
\frac{1}{8\pi}
\int_{S^2} {\rm d}\Omega_{\unit k}\>
\sum_{\alpha=1,2}
(\unit u_1\cdot\hat{\vecs\epsilon}_{\alpha}(\unit k))
(\unit u_2\cdot\hat{\vecs\epsilon}_{\alpha}(\unit k))
\left[1-e^{i\vec k\cdot\vec x_B}\right]
\left[1-e^{-i\vec k\cdot\vec x_C}\right]
\right\}
\,.
\label{e:Gamma12_EM}
\ee
\end{widetext}
If we ignore the contribution of the integrals involving
$e^{i\vec k\cdot\vec x_B}$,
$e^{-i\vec k\cdot\vec x_C}$,
and 
$e^{i\vec k\cdot(\vec x_B-\vec x_C)}$, assuming as we did for 
sound that we are working in the 
short-wavelength approximation, then
\be
\Gamma_{12}(\omega)\simeq
\frac{1}{8\pi}
\int_{S^2}{\rm d}^2\Omega_{\unit k} \>
\sum_{\alpha=1,2}
(\unit u_1\cdot \units\epsilon_\alpha(\unit k))
(\unit u_2\cdot \units\epsilon_\alpha(\unit k))\,,
\label{e:Gamma12_EM_approx}
\ee
which is the sky-averaged and polarization-averaged 
product of the inner products of $\unit u_1$ and 
$\unit u_2$ with the polarization vectors $\units\epsilon_\alpha(\unit k)$.

The above integral for the correlation function 
$\Gamma_{12}(\omega)$ can easily be 
evaluated in the coordinate system
shown in Fig.~\ref{f:EM-geom}.
In these coordinates, $\unit u_1=\unit z$ and 
$\unit u_2 = \sin\zeta\,\unit x+\cos\zeta\,\unit z$.
Using the expressions for the polarization vectors given in
(\ref{e:eps12}) it follows that
\be
\begin{aligned}
&\unit u_1 \cdot \units\epsilon_1(\unit k) = -\sin\theta\,,
\\
&\unit u_1 \cdot \units\epsilon_2(\unit k) = 0\,,
\\
&\unit u_2 \cdot \units\epsilon_1(\unit k)
= \sin\zeta\cos\theta\cos\phi - \cos\zeta\sin\theta\,,
\\
&\unit u_1 \cdot \units\epsilon_2(\unit k) =-\sin\zeta\sin\phi\,,
\end{aligned}
\ee
for which
\be
\begin{aligned}
\Gamma_{12}(\omega)
&\simeq
\frac{1}{8\pi}
\int_{S^2}{\rm d}^2\Omega_{\unit k} \>
\sum_{\alpha=1,2}
(\unit u_1\cdot \units\epsilon_\alpha(\unit k))
(\unit u_2\cdot \units\epsilon_\alpha(\unit k))
\\
&=\frac{1}{8\pi}
\int_0^{2\pi}{\rm d}\phi\>
\int_{-1}^1{\rm d}(\cos\theta)\>
(-\sin\theta)
\\
&\hspace{.75in}
\times(\sin\zeta\cos\theta\cos\phi - \cos\zeta\sin\theta)
\\
&=\frac{1}{4}\cos\zeta\int_{-1}^1{\rm d}x\>(1-x^2)
= \frac{1}{3}\cos\zeta\,.
\end{aligned}
\ee
Thus
\be
\chi(\zeta)\equiv \Gamma_{12}(\omega)
\simeq \frac{1}{3}\cos\zeta
\ee
and
\be
\langle r_1 r_2\rangle \simeq \frac{1}{3}\cos\zeta\,\langle E^2\rangle\,.
\ee
So the Hellings and Downs function for an isotropic, unpolarized stochastic
electromagnetic field is simply proportional to the cosine of the angle 
between the two receiving systems.

\section{Summary and discussion}
\label{s:summary}

In the preceeding two sections, we calculated 
Hellings-and-Downs-type functions $\chi(\zeta)$ for two simple scenarios: 
(i) omni-directional microphones in an isotropic stochastic sound field, and 
(ii) electric dipole antennas in an isotropic, unpolarized stochastic 
electromagnetic field.
The result for sound was trivial, $\chi(\zeta)={\rm const}$,
and in retrospect did not even require a calculation.
The result for the electromagnetic case was slightly more 
complicated, $\chi(\zeta)=\frac{1}{3}\cos(\zeta)$, 
as we had to take account of 
the polarization of the electromagnetic waves as well 
as the direction of the electric dipole antennas.
But the basic steps that we went through to obtain the
results were the same in both cases, and, in fact, 
can be abstracted to work for receivers in a general 
field, which we will denote here by $\Phi(t,\vec x)$:\cite{footnote5}
\begin{enumerate}

\item 
Write down the most general expression for the field
in terms of a Fourier expansion.
Let the Fourier coefficients be random variables 
whose expectation values encode the statistical properties 
of the field---e.g., isotropic, unpolarized, $\cdots$ .

\item 
Using the expectation values of the Fourier coefficients,
calculate $\langle \Phi^2(t, \vec x)\rangle$.
Use this expression to determine how the power in the 
field is distributed as a function of frequency 
\be
\langle \Phi^2(t, \vec x)\rangle
=\int_0^\infty {\rm d}\omega\>
\frac{{\rm d}\langle \Phi^2\rangle}{{\rm d}\omega}\,.
\ee
 
\item 
\label{s:convolution}
Write down the response $r_I(t)$ of receiver $I$ 
to the field $\Phi(t,\vec x)$.
For a linear receiving system, the response will 
take the form of a convolution:
\be
\begin{aligned}
r_I(t)
&= (R_I * \Phi)(t)
\\
&= \int {\rm d}\tau
\int {\rm d}^3 y\>
R_I(\tau, \vec y)
\Phi(t-\tau, \vec x_I - \vec y)\,.
\end{aligned}
\ee
For the simple examples we considered in
Sections~\ref{s:sound} and \ref{s:EM}, 
$R_I(\tau,\vec y)$ was proportional to a sum of
a product of delta functions like
$\delta(\tau)\delta^3(\vec y)$,
but that need not be the case in general.

\item Using the expectation values of the Fourier coefficients,
calculate the expected value of the correlated response
$\langle r_1(t) r_2(t)\rangle$ for any pair of receivers.
Use this expression to determine the correlation function
$\Gamma_{12}(\omega)$ defined by
\be
\langle r_1(t)r_2(t)\rangle
=
\int_0^\infty {\rm d}\omega\>
\frac{{\rm d}\langle \Phi^2\rangle}{{\rm d}\omega}
\Gamma_{12}(\omega)\,.
\ee

\item For fixed frequency $\omega$, the correlation 
$\Gamma_{12}(\omega)$ is, by definition, 
the value of the Hellings and 
Downs function evaluated for the relative 
configuration of the two receiving systems.
For example,
\be
\chi(\zeta) = \Gamma_{12}(\omega)
\ee
for the simple examples that we considered in 
Sections~\ref{s:sound} and \ref{s:EM}, where $\zeta$ 
is the angle between the two receiving systems, 
relative to an origin defined by the common 
microphone $A$ for sound, or the co-located dipole 
antennas $A$ and $A'$ for the electromagnetic field. 
For more complicated receivers, such as ground-based 
laser interferometers like LIGO, Virgo, etc., 
$\chi$ will be a function of {\em several} variables;
the separation vector between the vertices of 
the two interferometers $\vec s\equiv\vec x_1-\vec x_2$,
as well as the unit vectors $\unit u_1$, $\unit v_1$, and
$\unit u_2$, $\unit v_2$, which point along the arms of 
the two interferometers.

\end{enumerate}
The above five steps are generic and will work for any scenario.

We conclude this paper by stating {\em without proof} the expression 
for the actual gravitational-wave pulsar timing Hellings and Downs function:
\begin{widetext}
\be
\chi(\zeta) \equiv \frac{1}{8\pi}
\int_{S^2}{\rm d}^2\Omega_{\unit k} \>
\sum_{\alpha=+,\cross}
\frac{1}{2}
\left(\frac{\unit u_1\otimes\unit u_1}{1+\unit k\cdot\unit u_1}\right)
:\vecs\epsilon_\alpha(\unit k)
\,
\frac{1}{2}
\left(\frac{\unit u_2\otimes\unit u_2}{1+\unit k\cdot\unit u_2}\right)
:\vecs\epsilon_\alpha(\unit k)\,,
\label{e:HD_pulsartiming}
\ee
where 
\end{widetext}
\be
\unit u_I\otimes\unit u_I:\vecs\epsilon_\alpha(\unit k)
\equiv
\sum_{a=1}^3\sum_{b=1}^3 
u_I^a u_I^b\,\epsilon_{\alpha,ab}(\unit k)\,,
\quad I=\{1,2\}
\label{e:uueps}
\ee
and 
\be
\begin{aligned}
\vecs\epsilon_+(\unit k) 
&= \units\theta\otimes\units\theta
-\units\phi\otimes\units\phi\,,
\\
\vecs\epsilon_\cross(\unit k) 
&= \units\theta\otimes\units\phi
+\units\phi\otimes\units\theta\,,
\label{e:eps+x}
\end{aligned}
\ee
are the two gravitational-wave polarization tensors.
(Here $\unit u_1$ and $\unit u_2$ are unit vectors pointing from 
Earth to the two pulsars, and $\zeta$ is the 
angle between $\unit u_1$ and $\unit u_2$ as shown in 
Fig.~\ref{f:pulsar-geom}.)
The extra factors of 
$1/(1+\unit k\cdot \unit u_1)$ and
$1/(1+\unit k\cdot \unit u_2)$ 
that appear in (\ref{e:HD_pulsartiming})---as compared
to the analogous electromagnetic expression
(\ref{e:Gamma12_EM_approx})---come from the 
calculation of the timing residual response
of an Earth-pulsar baseline to the gravitational-wave
field, when integrating the metric pertubations
perturbations $h_{ab}(t,\vec x)$ along the photon world-line
from the pulsar to Earth.
This is a non-trivial example of the convolution described 
in Step~\ref{s:convolution} above, and the mathematical 
details needed to derive the precise form of (\ref{e:HD_pulsartiming}) 
are outside the scope of this paper.
(Readers who are interested in seeing a derivation of
(\ref{e:HD_pulsartiming}) are encouraged to consult
Ref.~\onlinecite{Anholm-et-al:2009}.)
But all in all, the pulsar timing Hellings and Downs function 
is just a sky-averaged and polarization-averaged product of two 
geometrical quantities, as is the case for any 
Hellings-and-Downs-type function.
It is now just a matter of doing the integrations, which we 
leave to the motivated reader.\cite{footnote6}
The final result should be proportional to (\ref{e:HDcurve}), which
has been normalized by an overall multiplicative factor of 3.


\begin{acknowledgments}
The content of this paper was originally presented by FAJ
during the student week of the 2011 IPTA summer school at 
the University of West Virginia.
JDR acknowledges support from NSF Awards PHY-1205585 and 
CREST HRD-1242090.
We also thank the referees for numerous suggestions that 
have improved the presentation of the paper.
\end{acknowledgments}

\appendix
\section{Details of the calculation for the pulsar timing
Hellings and Downs function}
\label{s:app}

Here we fill in some of the details of the integration
of the pulsar timing Hellings and Downs function 
(\ref{e:HD_pulsartiming}), following the
hints given in Endnote~\onlinecite{footnote6}.
The approach that we follow is based on similar 
presentations found in the
appendices of Refs.~\onlinecite{Anholm-et-al:2009, Mingarelli-et-al:2013,
Gair-et-al:2014}.

In the coordinate system shown in 
Fig.~\ref{f:pulsar-geom}, the two pulsars are located
in directions
$\unit u_1 = \unit z$ and
$\unit u_2 = \sin\zeta\,\unit x + \cos\zeta\,\unit z$, 
so that
\be
\begin{aligned}
&\unit k\cdot\unit u_1 
= \cos\theta\,,
\\
&\unit k\cdot\unit u_2 
= \cos\zeta\cos\theta + \sin\zeta\sin\theta\cos\phi\,.
\end{aligned}
\ee
Using the definition (\ref{e:eps+x}) 
of the gravitational-wave polarization tensors 
$\vecs\epsilon_\alpha(\unit k)$ with
$\units\theta$, $\units\phi$ defined in
(\ref{e:eps12}), it is fairly easy to show that
\begin{widetext}
\be
\begin{aligned}
\unit u_1\otimes\unit u_1:\vecs\epsilon_+(\unit k)
&=\sin^2\theta\,,
\\
\unit u_1\otimes\unit u_1:\vecs\epsilon_\cross(\unit k)
&=0\,,
\\
\unit u_2\otimes\unit u_2:\vecs\epsilon_+(\unit k)
&=(\sin\zeta\cos\theta\cos\phi -\cos\zeta\sin\theta)^2- \sin^2\zeta\sin^2\phi\,,
\\
\unit u_2\otimes\unit u_2:\vecs\epsilon_\cross(\unit k)
&=-2(\sin\zeta\cos\theta\cos\phi-\cos\zeta\sin\theta)\sin\zeta\sin\phi\,.
\end{aligned}
\ee
The quantities
\be
F_I^\alpha(\unit k)\equiv
\frac{1}{2}
\left(\frac{\unit u_I\otimes\unit u_I}{1+\unit k\cdot\unit u_I}\right)
:\vecs\epsilon_\alpha(\unit k)\,,
\quad
I=\{1,2\}\,,
\quad
\alpha=\{+,\cross\}\,,
\ee
which appear in (\ref{e:HD_pulsartiming}) are then given by
\be
\begin{aligned}
F_1^+(\unit k) &= \frac{1}{2}(1-\cos\theta)\,,
\\
F_1^\cross(\unit k) &= 0\,,
\\
F_2^+(\unit k) &= \frac{1}{2}
\left[(1-\cos\zeta\cos\theta-\sin\zeta\sin\theta\cos\phi)
-\frac{2\sin^2\zeta \sin^2\phi}
{1+\cos\zeta\cos\theta+\sin\zeta\sin\theta\cos\phi}
\right]\,,
\\
F_2^\cross(\unit k) &= -\frac{1}{2}
\left[
\frac{\sin^2\zeta \cos\theta\sin(2\phi) - \sin(2\zeta)\sin\theta\sin\phi}
{1+\cos\zeta\cos\theta+\sin\zeta\sin\theta\cos\phi}
\right]\,,
\end{aligned}
\ee
where for $F^+_2(\unit k)$ we cancelled the denominator with 
part of the numerator to isolate the complicated $\phi$-dependence.

In this reference frame, the pulsar timing Hellings and 
Downs function (\ref{e:HD_pulsartiming}) simplifies to
\be
\begin{aligned}
\chi(\zeta)
=\frac{1}{8\pi}\int_{S^2}{\rm d}^2\Omega_{\unit k}\>
F_1^+(\unit k)F_2^+(\unit k)
=\frac{1}{16\pi}\int_{-1}^1 {\rm d}x\>(1-x)I(x,\zeta)\,,
\end{aligned}
\ee
where $x\equiv\cos\theta$, and
\be
\begin{aligned}
I(x,\zeta)
&\equiv \int_0^{2\pi}{\rm d}\phi\> F_2^+(\unit k)
\\
&=\frac{1}{2}\int_0^{2\pi}{\rm d}\phi\> 
\left[(1-x\cos\zeta-\sqrt{1-x^2}\sin\zeta\cos\phi)
-\frac{2\sin^2\zeta \sin^2\phi}
{1+x\cos\zeta+\sqrt{1-x^2}\sin\zeta\cos\phi}
\right]\,.
\end{aligned}
\ee
The first part of the integral for $I(x,\zeta)$ is trivial:
\be
I_1(x,\zeta) 
\equiv 
\frac{1}{2}\int_0^{2\pi}{\rm d}\phi\> 
(1-x\cos\zeta-\sqrt{1-x^2}\sin\zeta\cos\phi)
\\
= \pi(1-x\cos\zeta)\,.
\ee
\end{widetext}
The second part can be be evaluated using contour 
integration,\cite{boas:2006} which we illustrate
below.
Making the usual substitutions $z=e^{i\phi}$,
$\cos\phi = \frac{1}{2}(z+z^{-1})$, etc., we obtain
\be
\begin{aligned}
I_2(x,\zeta) 
&\equiv -\sin^2\zeta
\int_0^{2\pi}{\rm d}\phi\> 
\frac{\sin^2\phi}
{1+x\cos\zeta+\sqrt{1-x^2}\sin\zeta\cos\phi}
\\
&=-\sin^2\zeta\oint_C {\rm d}z\>f(z)\,,
\end{aligned}
\ee
where 
\be
f(z) 
= \frac{i(z^2 - 1)^2}
{z^2\left[4z(1+x\cos\zeta)+2\sqrt{1-x^2}\sin\zeta(z^2+1)\right]}
\ee 
and $C$ is the unit circle in the complex $z$-plane.
The denominator of $f(z)$ can be factored using the
quadratic formula for the expression in square brackets:
\begin{multline}
4z(1+x\cos\zeta)+2\sqrt{1-x^2}\sin\zeta(z^2+1)
\\= 
2\sqrt{1-x^2}\sin\zeta(z-z_+)(z-z_-)\,,
\end{multline}
where
\begin{align}
z_+ 
\equiv 
-\sqrt{
\left(\frac{1\mp\cos\zeta}{1\pm\cos\zeta}\right)
\left(\frac{1\mp x}{1\pm x}\right)}\,,
\quad
z_- \equiv \frac{1}{z_+}\,.
\end{align}
In the above expression, the top signs correspond
to the region $-\cos\zeta\le x\le 1$ 
and the bottom signs to the region 
$-1\le x\le -\cos\zeta$.
One can show that for both of these regions, $z_+$ is 
inside the unit circle $C$ (i.e., $|z_+|\le 1$) 
and hence contributes to the 
contour integral, while $z_-$ is outside the unit circle
and does not contribute.
In addition, 
$z=0$ lies inside the unit circle and contributes
to the contour integral as a pole of order two.
Using the residue theorem\cite{boas:2006}
\be
\oint_C f(z)\,{\rm d}z = 2\pi i\sum_i {\rm Res}(f,z_i)\,,
\ee
with
\be
\begin{aligned}
&{\rm Res}(f,z_+) 
=\lim_{z\rightarrow z_+}
\left\{(z-z_+)f(z)\right\}
= \frac{i(z_+-z_-)}{2\sqrt{1-x^2}\sin\zeta}\,,
\\
&{\rm Res}(f,0)
=\lim_{z\rightarrow 0}
\left\{\frac{d}{dz}\left[z^2 f(z)\right]\right\}
= \frac{i(z_++z_-)}{2\sqrt{1-x^2}\sin\zeta}\,,
\end{aligned}
\ee
it follows that 
\be
\oint_C f(z)\,{\rm d}z 
= \frac{2\pi}{(1\pm x)(1\pm\cos\zeta)}\,,
\ee
for which
\be
I(x,\zeta) = \pi(1-x\cos\zeta) - 2\pi\frac{(1\mp\cos\zeta)}{(1\pm x)}\,.
\ee
It is now a relatively simple matter to the evaluate the 
integral over $x$ to obtain $\chi(\zeta)$:
\begin{widetext}
\be
\begin{aligned}
\chi(\zeta)
&= \frac{1}{16}
\Bigg\{\int_{-1}^1 {\rm d}x\>(1-x)(1-x\cos\zeta)
-2(1+\cos\zeta)\int_{-1}^{-\cos\zeta}{\rm d}x\
-2(1-\cos\zeta)\int_{-\cos\zeta}^{1}{\rm d}x\> \frac{(1-x)}{(1+x)}
\Bigg\}
\\
&= \frac{1}{16}
\Bigg\{2+\frac{2}{3}\cos\zeta
-2(1+\cos\zeta)(1-\cos\zeta)
-2(1-\cos\zeta)\left[2\ln\left(\frac{2}{1-\cos\zeta}\right) - (1+\cos\zeta)\right]
\Bigg\}
\\
&= \frac{1}{8}+\frac{1}{24}\cos\zeta
+\frac{1}{4}(1-\cos\zeta)\ln\left(\frac{1-\cos\zeta}{2}\right)
\\
&= 
\frac{1}{6}-\frac{1}{12}\left(\frac{1-\cos\zeta}{2}\right)
+\frac{1}{2}\left(\frac{1-\cos\zeta}{2}\right)\ln\left(\frac{1-\cos\zeta}{2}\right)\,.
\end{aligned}
\ee
\end{widetext}
Note that the above expression differs from (\ref{e:HDcurve}) by an overall normalization 
factor of $1/3$.
The normalization used in (\ref{e:HDcurve}) was chosen so that 
for zero angular separation, $\chi(\zeta)|_{\zeta=0}=1/2$ for two distinct pulsars.
This was purely an aesthetic choice, which does not change the angular 
dependence (i.e., shape) of the Hellings and Downs curve.



\begin{thebibliography}{10}

\bibitem{lorimer-kramer:2004}
D.~R. {Lorimer} and M.~{Kramer}, {\em {Handbook of Pulsar Astronomy}}.
\newblock Cambridge University Press, Cambridge, UK, Dec. 2004.

\bibitem{lorimer-LRR:2008}
D.~R. {Lorimer}, ``{Binary and Millisecond Pulsars},'' {\em Living Reviews in
  Relativity}, vol.~11 (8), pp.~1--90, Nov. 2008.

\bibitem{stairs-LRR:2003}
I.~H. {Stairs}, ``{Testing General Relativity with Pulsar Timing},'' {\em
  Living Reviews in Relativity}, vol.~6 (5), pp.~1--49, Sept. 2003.

\bibitem{weisberg-et-al:2010}
J.~M. {Weisberg}, D.~J. {Nice}, and J.~H. {Taylor}, ``{Timing Measurements of
  the Relativistic Binary Pulsar PSR B1913+16},'' {\em \apj}, vol.~722,
  pp.~1030--1034, Oct. 2010.

\bibitem{Hulse-Taylor:1975}
R.~A. Hulse and J.~H. Taylor, ``Discovery of a pulsar in a binary system,''
  {\em \apj}, vol.~195, no.~2, pp.~L51--L53, 1975.

\bibitem{estabrook-wahlquist:1975}
F.~B. {Estabrook} and H.~D. {Wahlquist}, ``{Response of Doppler spacecraft
  tracking to gravitational radiation},'' {\em General Relativity and
  Gravitation}, vol.~6, pp.~439--447, Oct. 1975.

\bibitem{sazhin:1978}
M.~V. {Sazhin}, ``{Opportunities for detecting ultralong gravitational
  waves},'' {\em \sovast}, vol.~22, pp.~36--38, Feb. 1978.

\bibitem{detweiler:1979}
S.~{Detweiler}, ``{Pulsar timing measurements and the search for gravitational
  waves},'' {\em \apj}, vol.~234, pp.~1100--1104, Dec. 1979.

\bibitem{footnote1}
{A pulsar timing model predicts the arrival times
of the pulses given values for the pulsar's spin frequency, 
frequency derivative, location on the sky, proper motion 
with respect to the solar system barycenter, its 
orbital parameters if the pulsar is in a binary, etc.\cite{lorimer-LRR:2008}
The values of these parameters are typically determined
by an iterative least-squares fitting procedure, which 
minimizes the root-mean-squared (rms) deviation of the 
resultant timing residuals.
Systematic errors in the timing model parameters can 
usually be identified by this iterative procedure, but 
unmodelled processes in the timing model will lead to errors 
in the timing residuals that cannot easily be removed.}

\bibitem{Foster-Backer:1990}
R.~S. {Foster} and D.~C. {Backer}, ``{Constructing a pulsar timing array},''
  {\em \apj}, vol.~361, pp.~300--308, Sept. 1990.

\bibitem{Hellings-Downs:1983}
R.~W. Hellings and G.~S. Downs, ``Upper limits on the istotropic gravitational
  radiation background from pulsar timing analysis,'' {\em \apj}, vol.~265,
  pp.~L39--L42, 1983.

\bibitem{Lee-et-al:2008}
K.~J. {Lee}, F.~A. {Jenet}, and R.~H. {Price}, ``{Pulsar Timing as a Probe of
  Non-Einsteinian Polarizations of Gravitational Waves},'' {\em \apj},
  vol.~685, pp.~1304--1319, Oct. 2008.

\bibitem{Mingarelli-et-al:2013}
C.~M.~F. {Mingarelli}, T.~{Sidery}, I.~{Mandel}, and A.~{Vecchio},
  ``{Characterizing gravitational wave stochastic background anisotropy with
  pulsar timing arrays},'' {\em \prd}, vol.~88, p.~062005(17), Sept. 2013.

\bibitem{Taylor-Gair:2013}
S.~R. {Taylor} and J.~R. {Gair}, ``{Searching for anisotropic
  gravitational-wave backgrounds using pulsar timing arrays},'' {\em \prd},
  vol.~88, p.~084001(25), Oct. 2013.

\bibitem{Gair-et-al:2014}
J.~{Gair}, J.~D. {Romano}, S.~{Taylor}, and C.~M.~F. {Mingarelli}, ``{Mapping
  gravitational-wave backgrounds using methods from CMB analysis: Application
  to pulsar timing arrays},'' {\em \prd}, vol.~90, p.~082001(44), Oct. 2014.

\bibitem{Shannon-et-al:2013}
R.~M. Shannon, V.~Ravi, W.~A. Coles, G.~Hobbs, M.~J. Keith, R.~N. Manchester,
  J.~S.~B. Wyithe, M.~Bailes, N.~D.~R. Bhat, S.~Burke-Spolaor, J.~Khoo,
  Y.~Levin, S.~Os{\l}owski, J.~M. Sarkissian, W.~van Straten, J.~P.~W.
  Verbiest, and J.-B. Wang, ``Gravitational-wave limits from pulsar timing
  constrain supermassive black hole evolution,'' {\em Science}, vol.~342,
  no.~6156, pp.~334--337, 2013.

\bibitem{footnote2}
{We are assuming here that the two pulsars---even for the case 
$\zeta=0$---are distinct (i.e., they do not occupy the same physical
location in space).
If we consider the same pulsar, as would be the case for an 
autocorrelation calculation, then the right-hand-side of (\ref{e:HDcurve}) 
should have an extra term equal to $\delta(\zeta)/2$.}

\bibitem{footnote3}
{This statement is a generalization (to fields) of 
the mathematical result that the Fourier transform of the 
probability distribution $p(x)$ for a random variable $x$ 
(i.e., the so-called {\em characteristic function} of the 
random variable)
can  be written as a power series with coefficients given by the expectation values 
$\langle x^k\rangle$ for $k=1,2,\cdots$.\cite{helstrom:1968}
Thus, the expectation values $\langle x^k\rangle$ for $k=1,2,\cdots$
completely determine the probability distribution $p(x)$ and hence
the statistical properties of the random variable $x$.}

\bibitem{helstrom:1968}
C.~W. {Helstrom}, {\em {Statistical Theory of Signal Detection}}.
\newblock Pergamon Press, Oxford, United Kingdom, 1968.

\bibitem{footnote4}
{Recall:
$\del^2 = \left(\frac{\partial^2}{\partial x^2} + \frac{\partial^2}{\partial y^2} 
+ \frac{\partial^2}{\partial z^2}\right)$ in Cartesian coordinates $(x,y,z)$.}

\bibitem{footnote5}
{The field might actually be a tensor field, like the 
gravitational-wave field $h_{ab}(t, \vec x)$, and hence should
have tensor indices in general.
But for simplicity, we will ignore that complication here.}

\bibitem{Anholm-et-al:2009}
M.~{Anholm}, S.~{Ballmer}, J.~D.~E. {Creighton}, L.~R. {Price}, and
  X.~{Siemens}, ``{Optimal strategies for gravitational wave stochastic
  background searches in pulsar timing data},'' {\em \prd}, vol.~79,
  p.~084030(19), Apr. 2009.

\bibitem{footnote6}
{Hint: Work in the coordinate system shown 
in Fig.~\ref{f:pulsar-geom} with the Earth at the 
origin and the two pulsars located along the $z$-axis
and in the $xz$-plane, respectively.
Evaluate (\ref{e:uueps}) in this frame using 
(\ref{e:eps+x}) and the expressions for
$\units\theta$, $\units\phi$ given in (\ref{e:eps12}).
Finally, use contour integration to do the integral over the azimuthal 
angle $\phi$.
It is a long calculation, but worth the effort.
If you have trouble completing the calculation, see
Appendix~\ref{s:app} for more details.\label{fn:hint}}

\bibitem{boas:2006}
M.~L. {Boas}, {\em {Mathematical methods in the physical sciences}}.
\newblock John Wiley \& Sons, Inc., Hoboken, NJ, 2006.

\end{thebibliography}


\end{document}